\documentclass[twocolumn,a4paper,10pt,groupedaddress]{revtex4-1}
\usepackage[utf8]{inputenc}
\usepackage{color}
\usepackage{graphicx}
\usepackage{amsmath}
\usepackage{hyperref}
\graphicspath{{graphics/}}

\begin{document}
\title{Efficient simulation of the random-cluster model}
\author{Eren Metin El\c{c}i}
\author{Martin Weigel}
\affiliation{Applied Mathematics Research Centre, Coventry University, Coventry, CV1 5FB, United Kingdom}
\affiliation{Institut f\"ur Physik, Johannes Gutenberg-Universit\"at Mainz,
  Staudinger Weg 7, D-55099 Mainz, Germany}
\date{\today}
\begin{abstract}
  The simulation of spin models close to critical points of continuous phase
  transitions is heavily impeded by the occurrence of critical slowing down. A number
  of cluster algorithms, usually based on the Fortuin-Kasteleyn representation of the
  Potts model, and suitable generalizations for continuous-spin models have been used
  to increase simulation efficiency. The first algorithm making use of this
  representation, suggested by Sweeny in 1983, has not found widespread adoption due
  to problems in its implementation.  However, it has been recently shown that it is
  indeed more efficient in reducing critical slowing down than the more well-known
  algorithm due to Swendsen and Wang. Here, we present an efficient implementation of
  Sweeny's approach for the random-cluster model using recent algorithmic advances in
  dynamic connectivity algorithms.
\end{abstract}
\maketitle

\section{Introduction}

In the vicinity of a continuous phase transition, particle or spin systems of
statistical mechanics develop extended spatial correlations signaling the onset of
long-range translational order through spontaneous symmetry breaking. It has been
realized early on that these phenomena suggest a description of the ordering in
geometrical terms, using analogies to the percolation transition
\cite{fisher:67}. While Fisher's droplet model initially considered simple clusters
of like spins (geometrical clusters) as the relevant quantities, it was only
gradually realized in the 1980s that the relevant collective degrees of freedom, or
``physical clusters'', are indeed of a different nature, and they can be constructed
by breaking the geometric clusters up following a suitable stochastic prescription
activating bonds within clusters with probability $p = 1-\exp(-J/k_B T) < 1$
\cite{coniglio:80a}. Initially, this breakup was only applied to one spin species (of
the Ising model, say), leading to inconsistencies in the symmetric high-temperature
phase. It was in an independent line of thought by Fortuin and Kasteleyn
\cite{fortuin:72a,*fortuin:72b,*fortuin:72c} that an equivalence of the partition
function of the $q$-state Potts model \cite{wu:82a} to a correlated bond-percolation
problem known as the random-cluster model with partition function
\begin{equation}
  \label{eq:random_cluster}
  Z_\mathrm{RC} = \sum_{{\cal G}'\subseteq{\cal G}}v^{b({\cal G}')}q^{n({\cal G}')}
\end{equation}
was established. Here, ${\cal G}'$ denotes the set of $b({\cal G}')$ activated edges
on a lattice graph ${\cal G}$, resulting in $n({\cal G}')$ connected components, and
the bond weight $v = p/(1-p)$. These results led Hu \cite{hu:84a} to generalize the
above bond activation probability to {\em all\/} geometric clusters irrespective of
their orientation. As a consequence, the right choice of physical droplets is now
well understood, and the equivalence of their percolation properties and thermal
quantities has been explicitly checked \cite{demeo:90}.

The growth without bounds of static correlations in critical systems is accompanied
in the time domain by a divergence of relaxation times known as {\em critical slowing
  down\/} \cite{hohenberg:77a}. While this is a physical phenomenon connected, for
instance, to the effect of critical opalescence, it is also of direct relevance for
the pseudo-dynamics in Monte Carlo simulations of near-critical systems. As this
leads to an asymptotic inefficiency of Markov chain Monte Carlo in producing
independent samples, an improved understanding of spatial correlations was hoped to
translate into suitable non-local updating procedures allowing to precisely study
near-critical systems. Initial attempts in this direction, such as variants of the
multi-grid approach \cite{schmidt:83,goodman:86}, were based on renormalization group
ideas, and turned out to be only moderately successful. The first Monte Carlo
algorithm based on the concept of physical clusters was suggested by Sweeny in 1983
\cite{sweeny:83}. He considered a direct simulation of the bond variables of
Eq.~\eqref{eq:random_cluster}, randomly suggesting state switches from active to
inactive and vice versa. As the relevant Boltzmann weight depends on the number
$n({\cal G}')$ of connected components resulting from a given bond configuration,
calculating the acceptance probability of the bond moves needs up-to-date information
about cluster connectivity. Hence, a single update might require the expensive
traversal of large (possibly spanning) clusters, potentially destroying the
computational advantage of an accelerated decorrelation of configurations through
{\em computational\/} critical slowing down \cite{deng:10}.

An alternative suggestion made by Swendsen and Wang \cite{swendsen-wang:87a} works
directly on the spin configuration, freezing bonds between like spins with
probability $p = 1-\exp(-J/k_B T)$ and independently flipping the resulting spin
clusters. Instead of working in the graph language alone, this series of alternating
updates of spin and bond variables corresponds to a Markov chain in an augmented
state space \cite{edwards:88a}. The resulting algorithm (with its many variants
including, for instance, the single-cluster version \cite{wolff:89a}) is rather
straightforward to implement and turns out to be very efficient in beating critical
slowing down, reducing the dynamical critical exponent, e.g., of the 2D Ising model
from $z\approx 2$ for local spin flips to $z \approx 0.2$ \cite{garoni:11}. Owing to
this success of the Swendsen-Wang algorithm and related techniques as well as the
delicacies of maintaining up-to-date connectivity information, Sweeny's approach was
not used by many researchers. Also, its reduction of critical slowing down was not
precisely investigated until, about 20 years after the original work, it was claimed
that a variant of the single-bond algorithm was completely free of critical slowing
down \cite{gliozzi:02}. Although this was later shown to be incorrect \cite{wang:02},
it was not until recently that its dynamical critical behavior was investigated in
more detail \cite{qian:05a,deng:07}, revealing the surprising feature of {\em
  critical speeding up\/}, i.e.\ $z < 0$, for certain ranges of $q$ alongside
generally smaller dynamical critical exponents than those found for the Swendsen-Wang
dynamics.

Besides it being a very elegant and direct sampling procedure for the weights of
Eq.~\eqref{eq:random_cluster}, another favorable feature of Sweeny's approach is its
general applicability to arbitrary values of $q$: while the Potts model is only
defined for integer $q = 1$, $2$, $3$, $\ldots$, the random cluster model of
Eq.~\eqref{eq:random_cluster} is meaningful for any real value $q\ge 0$, serving as
an analytic continuation of the Potts model to real $q$ \cite{grimmett:book}. The
Swendsen-Wang algorithm, originally working with a joint spin and bond
representation meaningful only for integer $q$, can be generalized to non-integer
$q\ge 1$ \cite{chayes:98a}. The bond algorithm, however, is the only approach for $0
< q< 1$. This fact has prompted a number of researchers to use Sweeny's approach to
probe the $q< 1$ regime, for instance to study fractal properties of the cluster
structure \cite{qian:05,zatepelin:10,gliozzi:10}. The main obstacle to a more
widespread adoption, however, has been the problem of expensive connectivity checks:
inserting an edge might join two previously unconnected clusters, deleting a bond can
lead to cluster fragmentation.  A naive approach without additional data structures
appears to require the tracing out of one (or two) randomly chosen cluster(s) to
check for connectivity. As the average cluster size scales proportional to
$L^{\gamma/\nu}$ \cite{stauffer:book} and $\gamma/\nu \ge 1.75$ for the
random-cluster model, the cost of a full lattice sweep is almost squared as compared
to single spin flips or Swendsen-Wang. In his paper, Sweeny had suggested a specific
solution for the case of two-dimensional lattices, replacing the traversal of
clusters with a tracking of boundary loops on the medial lattice
\cite{sweeny:83,deng:10}. Irrespective of space dimension, a pair of interleaved
breadth-first searches starting from both ends of the bond currently examined can
also dramatically improve the situation \cite{weigel:10d,deng:10}. While these
connectivity algorithms still exhibit power-law scaling with the size of the system,
fully dynamic connectivity algorithms, where edge insertions and removals can be
performed in amortized times at most (poly){\em logarithmic\/} in the system size,
are known in computer science \cite{henzinger:99,holm:01}. Here, we compare a number
of different implementations of Sweeny's algorithm for simulations of the
random-cluster model to each other as well as to the Chayes-Machta-Swendsen-Wang
dynamics \cite{swendsen-wang:87a,chayes:98a}. The combination of a polylogarithmic
dynamic connectivity algorithm and Sweeny's single-bond approach is shown to be the
more efficient way, asymptotically, to simulate the random-cluster model at
criticality.

The rest of the paper is organized as follows. In Sec.\ \ref{sec:algorithms}, we
introduce Sweeny's algorithm in more detail and describe the three different variants
of connectivity checks implemented here: breadth-first search, union-and-find, and
dynamic connectivities. Section \ref{sec:results} contains an in-depth comparison of
the scaling of properties of these approaches as compared to the
Chayes-Machta-Swendsen-Wang dynamics in terms of simulation as well as computer time
for the case of simulations on the square lattice. Finally, Sec.\ \ref{sec:concl}
contains our conclusions.

\section{Model and algorithms\label{sec:algorithms}}

The random-cluster model (RCM) assigns weights to (spanning) sub-graph configurations
${\cal G}'$, i.e., subsets of activated edges and the complete set of vertices, of the underlying graph
${\cal G}$ according to \cite{grimmett:book}
\begin{equation}
  w_\mathrm{RC}({\cal G}') = q^{n({\cal G}')} v^{b({\cal G}')},
  \label{eq:cluster_weight}
\end{equation}
leading to the partition sum of Eq.~\eqref{eq:random_cluster}. For integer values of
the cluster weight $q$, the partition function \eqref{eq:random_cluster} is identical
\cite{fortuin:72a} to that of the $q$-state Potts model with Hamiltonian
\begin{equation}
  \label{eq:potts_hamiltonian}
  {\cal H} = -J \sum_{b\in {\cal G}} \delta_{\sigma_i,\sigma_j},
\end{equation}
where $b=(i,j)$ is an edge in the graph ${\cal G}$, and $\sigma_i \in
\{1,\ldots,q\}$. For the purposes of this study, we will restrict ourselves to graphs
in two dimensions (2D), namely compact $L\times L$ regions of the square lattice,
applying periodic boundary conditions. For this case, the ordering transition of the
Potts model occurs at the coupling $J/k_B T = \ln(1+\sqrt{q})$, corresponding to the
critical bond weight $v_c = \sqrt{q}$ in \eqref{eq:random_cluster}. This transition
is continuous for $q\le 4$ and first-order for $q > 4$ \cite{wu:82a}.

\subsection{Sweeny's algorithm\label{sec:sweeny}}

Starting from the results of Fortuin and Kasteleyn \cite{fortuin:72a}, Sweeny
suggested to directly sample bond configurations of the RCM according to the weight
\eqref{eq:cluster_weight}. For any sub-graph ${\cal G}'$, the basic update operation
is then given by the deletion of an occupied edge or the insertion of an unoccupied
edge. According to Eq.~\eqref{eq:cluster_weight}, the corresponding transition
probabilities depend on the changes $\Delta b$ of the number of active edges and
$\Delta n$ of the number of connected components or clusters. While $\Delta b$ is
trivially determined to equal $+1$ for edge insertion and $-1$ for edge removal,
respectively, the change in cluster number depends on whether a chosen inactive edge
is {\em internal\/} to one cluster ($\Delta n = 0$) or, instead, it is {\em
  external\/} and hence amalgamates two existing clusters if activated ($\Delta n =
-1$). Likewise, removing an edge might lead to $\Delta n = 0$ or $\Delta n = +1$,
depending on whether an alternative path exists connecting the end points of the
removed edge. The construction and implementation of data structures supporting the
efficient calculation of $\Delta n$ constitutes the intricacy of Sweeny's algorithm
and the focus of the present work.

Importance sampling for the weight \eqref{eq:cluster_weight} can be constructed along
well-known lines, the most common choices being the heat-bath and Metropolis schemes
\cite{binder:book2}. In both cases, a bond is randomly and uniformly selected from
the graph and a ``flip'' of its occupation state from inactive to active or vice
versa is proposed. The heat-bath acceptance ratio, used in the original approach of
Sweeny \cite{sweeny:83} is then given by
\begin{equation}
  p_\mathrm{acc}^\mathrm{HB}(\Delta b, \Delta n) = \frac{q^{\Delta n}v^{\Delta b}}{1+q^{\Delta
      n}v^{\Delta b}}.
  \label{eq:heatbath}
\end{equation}
For Metropolis-Hastings, on the other hand, we have
\begin{equation}
   p_\mathrm{acc}^\mathrm{MH}(\Delta b, \Delta n) = \min(1,q^{\Delta n}v^{\Delta b}).
   \label{eq:metropolis}
\end{equation}
It is easily seen that
\[
\frac{1}{2} \le \frac{p_\mathrm{acc}^\mathrm{HB}}{p_\mathrm{acc}^\mathrm{MH}} =
\frac{\max(1,q^{\Delta n}v^{\Delta b})}{1+q^{\Delta n}v^{\Delta b}} < 1.
\]
Depending on $q$ and $v$, we hence expect up to twice larger acceptance rates for the
Metropolis variant. At criticality, $v=\sqrt{q}$, the minimal ratio
$p_\mathrm{acc}^\mathrm{HB}/p_\mathrm{acc}^\mathrm{MH} = 1/2$ is reached in the
percolation limit $q=1$. In contrast to Ref.~\cite{sweeny:83}, our numerical
experiments concentrate on Metropolis acceptance.

As, depending on the data structures used, the determination of the change $\Delta n$
in cluster number is the most expensive operation, it is economic to only determine
$\Delta n$ if it is actually required for the update. In an update attempt, one draws
a random number uniformly in $[0,1[$; if $r\le p_\mathrm{acc}$ the move is accepted,
otherwise it is rejected. Given that
\[
r\le \min_{\Delta n}\,p_\mathrm{acc}(\Delta b, \Delta n),
\]
where $\Delta n \in \{0,\,-1\}$ for insertion and $\Delta n \in \{0,\,+1\}$ for
deletion, respectively, the move can be {\em unconditionally accepted}
\cite{gliozzi:02}. Conversely, for
\[
r >  \max_{\Delta n}\,p_\mathrm{acc}(\Delta b, \Delta n)
\] {\em unconditional rejection} occurs. At criticality, $v=\sqrt{q}$, this results
in a fraction
\[
\min(\sqrt{q},1/\sqrt{q})
\]
of moves which can be unconditionally accepted or rejected under the Metropolis
dynamics \footnote{Note that in the Metropolis dynamics there are no unconditional
  rejections, while unconditional acceptance and rejection occur at equal rates for
  heat-bath rates.}. Likewise, for the heat-bath rate \eqref{eq:heatbath}, a fraction
\[
 \frac{2}{1+\sqrt{q}}\,\min(1,\sqrt{q})
\]
of move attempts can be decided without actually working out $\Delta n$. Note that,
in both cases, these fractions tend to unity as $q\to 1$ which is a result of the
cluster weight \eqref{eq:cluster_weight} becoming independent of cluster number in
the uncorrelated percolation limit. Connectivity checks are hence {\em never\/}
required there.

\subsection{Connectivity algorithms\label{sec:connectivity}}

The main complication for an efficient implementation of the bond algorithm is to
maintain the full connectivity information of the current sub-graph.  Consider a flip
attempt on a random edge; it can be currently in the active or inactive (inserted or
deleted) state. For each of these cases, one needs to distinguish internal from
external edges, such that independent paths connecting the two end points either
exist (internal edge) or are absent (external edge). This leads to the four cases of
internal/external insertions/deletions, each of which can exhibit rather different
runtime scaling behavior depending on the chosen implementation.

\subsubsection{Breadth-first search\label{sec:bfs}}

\begin{table}[bt]
  \centering
  \caption{Asymptotic run-time scaling at criticality of the elementary operations of insertion or
    deletion of internal or external edges, respectively, using sequential
    breadth-first search (SBFS), interleaved BFS (IBFS), union-and-find (UF) or the
    fully dynamic connectivity algorithm (DC) as a function of the linear system size
    $L$.}
  \label{tab:scaling}
  \begin{ruledtabular}
    \begin{tabular}{lllll}
      \multicolumn{1}{c}{move} & \multicolumn{1}{c}{SBFS} &  \multicolumn{1}{c}{IBFS} &
      \multicolumn{1}{c}{UF} & \multicolumn{1}{c}{DC} \\  \hline
      internal insertion & $L^{d_F-x_2}$          & $L^{d_F-x_2}$ & const.         & $\log^2 L$ \\
      external insertion & $L^{\gamma/\nu}$& $L^{d_F-x_2}$ & const.         &$\log^2 L$\\
      internal deletion  & $L^{d_F-x_2}$          & $L^{d_F-x_2}$ & $L^{d_F-x_2}$   &$\log^2 L$\\
      external deletion  & $L^{\gamma/\nu}$& $L^{d_F-x_2}$ & $L^{\gamma/\nu}$ &$\log^2 L$\\ \hline
      dominant & $L^{\gamma/\nu}$ &  $L^{d_F-x_2}$ & $L^{\gamma/\nu}$ &$\log^2 L$\\
    \end{tabular}
  \end{ruledtabular}
\end{table}

The simplest approach to the connectivity problem is to not maintain any state
information about clusters and determine the value of $\Delta n$ for each individual
bond move from direct searches in the graph structure around the current edge $e =
(i,j)$. Such traversals are most naturally implemented as breadth-first search (BFS)
\cite{cormen:09} starting from one of the end-points, say $i$, while not being
allowed to cross the edge $e$.  In case of an external edge, the cluster attached to
$i$ needs to be fully traversed. For an internal edge, on the other hand, the search
starting at $i$ terminates once it arrives at $j$, having found an alternative path
connecting $i$ and $j$.  Instead of the BFS one could also use a depth-first search
(DFS) to achieve the same result \cite{cormen:09}. We found essentially no
differences in the run-time behavior of both variants, however, and hence did not
consider this possibility in more detail. To determine the asymptotic run-time at
criticality of these operations, we note that the average number of clusters 
with mass $s$ per lattice site is \cite{stauffer:book}
\begin{equation}
  n_s \sim s^{-\tau} e^{-c s},
\end{equation}
where $\tau$ is the cluster-size or Fisher exponent. A randomly picked site will
therefore, on average, belong to a cluster of size
\[
M_2 = \sum_s s^2 n_s \sim \int s^{2-\tau} e^{-cs} \propto c^{\tau-3} \sim
L^{(\tau-3)/\sigma\nu},
\]
where the last identity follows from $c\sim |p-p_c|^{1/\sigma}$ and the standard
finite-size scaling ansatz $|p-p_c| \sim L^{1/\nu}$. Since $(3-\tau)/\sigma =
\gamma$, we arrive at a typical cluster size $M_2 \sim L^{\gamma/\nu}$
\cite{stauffer:book}. For operations on external edges, we therefore expect an
asymptotic scaling of run-times $\sim L^{\gamma/\nu}$. For internal edges, on the
other hand, the relevant effort corresponds to the total number of visited sites of a
breadth-first search starting from site $i$ until it reaches $j$. In this case, the
number of shells $\ell$ in the BFS at termination is just the shortest path between
$i$ and $j$. As shown by Grassberger \cite{grassberger:92a,grassberger:99}, for
bond percolation, the
probability of two nearby points on the lattice to be connected by a shortest path of
length $\ell$ is $p(\ell) \sim l^{-\psi_\ell}\mathcal{L}(\ell/L^{d_\mathrm{min}})$, where
\[
\psi_\ell = 1+\frac{2\beta}{\nu d_\mathrm{min}}+\frac{g_1}{d_\mathrm{min}}.
\]
Here, $d_\mathrm{min}$ is the shortest-path fractal dimension \cite{grassberger:92}
and $g_1$ is the scaling exponent related to the density of growth sites
\cite{grassberger:99}. Ziff \cite{ziff:99} demonstrated that $g_1 = x_2-2\beta/\nu$,
where $x_2$ is the two-arm scaling exponent \cite{cardy:98,smirnov:01}. Hence
$\psi_\ell = 1+x_2/d_\mathrm{min}$. As a result, the average length of shortest
path between nearby points exhibits system-size scaling according to
\begin{equation}
  \label{eq:path_scaling}
  \langle \ell\rangle \sim L^{d_\mathrm{min}-x_2}.
\end{equation}
The number of sites touched by a BFS from $i$ to $j$ separated by a shortest path of
length $\ell$ is expected to be $\ell^{\hat{d}}$, where $\hat{d} =
d_F/d_\mathrm{min}$ is known as spreading dimension \cite{grassberger:92}. Here, $d_F
= d-\beta/\nu$ denotes the fractal dimension of the percolating cluster. Hence, the
average number of sites touched by the BFS for an internal edge is
\begin{equation}
  \label{eq:sbfs_Scaling}
  \langle \ell^{\hat{d}}\rangle \sim L^{d_F-x_2}.
\end{equation}
Note that, while $d_F$ and $x_2$ are exactly known
\cite{nienhuis:domb,stauffer:book}, this is not the case for $d_\mathrm{min}$
\cite{deng:10,zhou:12}.

The asymptotic run-time scaling of the Sweeny update using sequential BFS (SBFS)
hence depends on the fractions of internal and external edges encountered. These are
found to be asymptotically $L$ independent numbers $0<r_\mathrm{int},r_\mathrm{ext} <
1$ which, however, vary with $q$ \cite{gyure:92,elci:prep}. Comparing the scaling
exponents for the operations on internal and external edges, it is found that
$\gamma/\nu > d_F-x_2$ for the whole range $0\le q\le 4$, cf.\ the data compiled in
Table \ref{tab:exp_runtime}. As a consequence, the stronger scaling $\sim
L^{\gamma/\nu}$ of the operations on external edges will always dominate the running
time in the limit of large system sizes.

An improvement suggested in Refs.~\cite{doktor,weigel:10d,deng:10} concerns the quasi
simultaneous execution of both BFSs. In practice, no hardware-level parallelism is
needed here and, instead, sites are removed from the BFS queues of the two searches
in an alternating fashion, effectively leading to an {\em interleaved\/} structure of
the cluster traversals. To understand the benefit of this modification, consider the
insertion of an edge $e$. If it is external, $i$ and $j$ belong to separate clusters
${\cal C}_1$ and ${\cal C}_2$ after the deletion of $e$. In this case, the searches
terminate as soon as the smaller of the two clusters has been exhausted, i.e., after
$C_{2,\mathrm{min}} \equiv \min(|{\cal C}_1|, |{\cal C}_2|)$ steps. Deng {\em et
  al.\/} \cite{deng:10} have shown that, at criticality, this minimum scales as
$L^{d_F-x_2}$, i.e., with the exponent already found above for operations on internal
edges in SBFS. For the case of an internal edge, the interleaved searches terminate
as soon as they meet each other. As argued in Ref.~\cite{deng:10} this time again
exhibits the same run-time scaling $\sim L^{d_F-x_2}$ which is hence the relevant
asymptotic behavior of the critical bond algorithm using interleaved BFS (IBFS). In
Table \ref{tab:scaling} we compare the run-time scaling of the elementary operations
between the different implementations of connectivity checks considered here.

\subsubsection{Union-and-find}

\begin{figure}
  \begin{tabular}{@{}lll@{}}
    \includegraphics[scale=0.81]{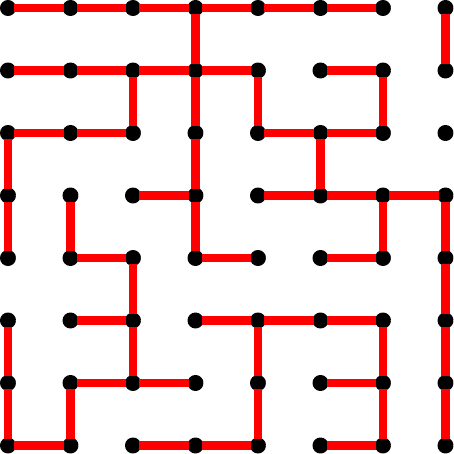} &\hspace{0.01\textwidth} &
    \includegraphics[scale=0.81]{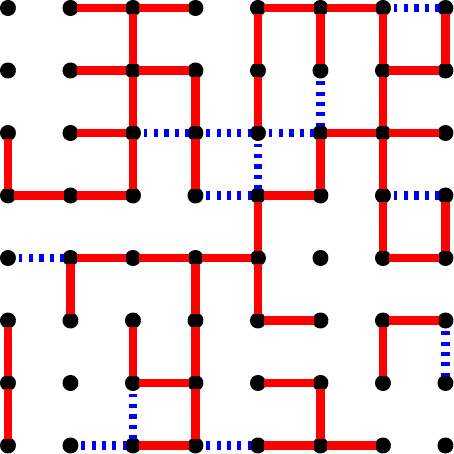}   \\ \vspace{0.0\textwidth} \\
    \includegraphics[scale=0.81]{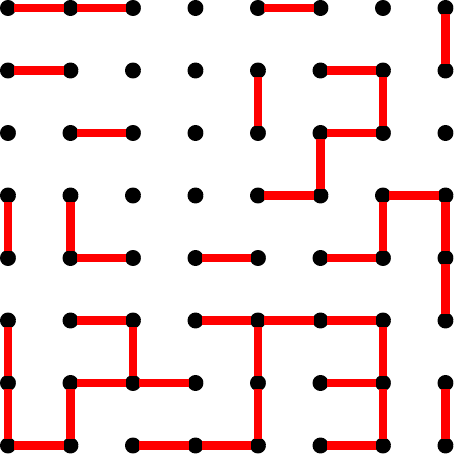} &\hspace{0.0\textwidth} &
    \includegraphics[scale=0.81]{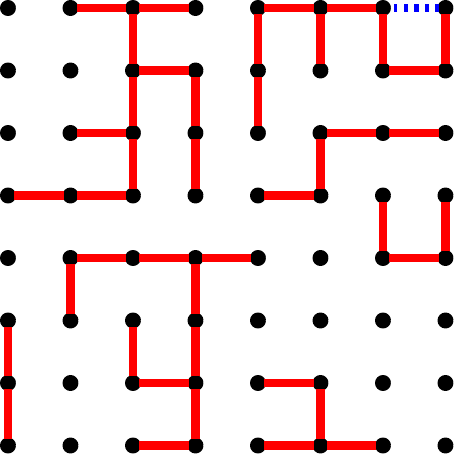}   \\ \end{tabular}
  \caption{\label{fig:edge_vis}
    (Color online) Equilibrium bond configurations for a $L=8$ system at criticality, $v_c = \sqrt{q}$.
    The left (right) column shows a configuration for $q = 0.0005$ ($q = 2$).
    The first (second) row corresponds to the level $0$ ($1$) graph.
    Here, tree edges are indicated in red, solid lines, while non-tree edges are
    drawn with blue, dotted lines.
}
\end{figure}

Sequential amalgamations of clusters through the addition of bonds can be handled
efficiently using tree-based data structures under a paradigm known as
union-and-find. This is traditionally applied to set partitioning \cite{cormen:09},
but has also been used in lattice models for highly efficient simulations of the bond
percolation problem \cite{newman:01a}. Each cluster is represented as a directed tree
of nodes with pointers to their parent nodes; the root corresponds to a designated
site representing the cluster as a whole. In this data structure, connectivity
queries are answered by path traversal to the root sites, such that two sites are
connected if and only if they have the same root. Using path compression
\cite{cormen:09}, where (most of) the node pointers directly link to the cluster
root, as well as a balancing heuristic that attaches the smaller cluster to the root
of the bigger in case of cluster fusion, allows to perform the connectivity check
with a worst-time scaling practically indistinguishable from a constant
\cite{tarjan:75}.

\begin{table*}[tb!]
  \caption{\label{tab:z_ints}
    Estimated dynamical critical exponents $z_{\mathrm{int},{\cal O}}$ for the
    two-dimensional RCM at criticality and ${\cal O} = \mathcal{S}_2$ as well as ${\cal O} =
    \mathcal{N}$ for a range of $q$ values as compared to results reported in
    Refs.~\cite{deng:07,deng:07a,garoni:11}. The values shown for
    $z_{\mathrm{int},\mathcal{S}_2}^\mathrm{SW}$ for the Swendsen-Wang-Chayes-Machta
    algorithm are actually related to another observable $\mathcal{E}$, but it was
    reported in Refs.~\cite{deng:07a,garoni:11} that the observables $\mathcal{E}'$,
    ${\cal N}$ and ${\cal S}_2$ share the same dynamical critical exponent for this
    algorithm.
  }
  \begin{ruledtabular}
    \renewcommand\arraystretch{1.4}
      \begin{tabular}{crccrccc} 
       $q$  & \multicolumn{1}{c}{$z_{\mathrm{int},\mathcal{S}_2}$}
          &$z_{\mathrm{int},\mathcal{S}_2}$ \cite{deng:07}
          &$z_{\mathrm{int},\mathcal{S}_2}^\mathrm{SW}$ \cite{deng:07a,garoni:11} &
          \multicolumn{1}{c}{$z_{\mathrm{int},\mathcal{N}}$} &$\alpha/\nu$
          &$z_{\mathrm{int},\mathcal{N}}$ \cite{deng:07} &
          $z_{\mathrm{int},\mathcal{S}_2}^\mathrm{SW} -
                                 z_{\mathrm{int},\mathcal{S}_2}$
                                  \\ \hline
      0.0005    & $-1.12(1)$         &-1.23  &    ---    & $0.01(1)   $     & -1.958    &   0       &--- \\
      0.005     & $-1.09(1) $         &-1.21  &    ---    & $0.01(1)    $ & -1.868    &   0       &--- \\
      0.05      & $-1.04(1) $         &  -1.12&    ---    & $0.01(1)    $ & -1.601    &   0       &--- \\
      0.2       & $-0.86(1) $         &  -1.01&    ---    & $-0.01(1)  $ &-1.247     &   0       &---\\
      0.5       & $-0.63(1) $         &-0.71  &    ---    & $0.00(1)    $     & -0.878    &   0       &---\\
      1.0       & $-0.33(1) $         &-0.32  &    ---    & $0.00(1)   $ & -0.500    &   0       &---\\
      1.5       & $-0.11(2) $         &-0.16  &  0& $0.06(2)    $ & -0.227    &   0       &\\
      2.0       & $0.03(3)  $         &-0.08  &  0.143(3) & $0.13(2)    $ & 0
                                  &   0           (log)&0.11\\
      3.0       & $0.44(4)  $         &0.41   &  0.497(3) & $0.45(4)    $ &
                                  0.400     &   0.45(1) &0.06\\
      4.0       & $0.75(6)  $         &---    &  0.910(5) & $0.73(6)    $ & 1
                                  &   ---     &0.16
    \end{tabular}
  \end{ruledtabular}
\end{table*}

Using this data structure for an implementation of the bond algorithm for the RCM
\cite{elci:11}, edge insertion requires a connectivity check. If the edge is
identified as internal, the cluster structure remains unchanged on its insertion
which hence can be performed in constant time. For an external edge, insertion is
realized through the attachment of the cluster root of the smaller cluster to the
bigger which is, again, a constant-time operation. For the deletion of an edge
$e=(i,j)$, the information about alternate paths between $i$ and $j$ is not directly
contained in the data structure. We hence use interleaved BFS to detect such paths,
with a computational effort asymptotically proportional to $L^{d_F-x_2}$. For an
internal edge, this completes the deletion. For the case of an external edge, leading
to fragmentation of the original cluster, a complete re-labeling of both new clusters
is required, however, resulting in a total scaling of $\sim L^{\gamma/\nu}$ for this
step, cf.\ Table \ref{tab:scaling}.

The total effective runtime of a bond simulation with union-and-find data structure
depends on the frequency of the individual operation types. The average number of
active bonds at criticality can be worked out from results for the square-lattice
Potts model, where the critical internal energy density is found to be $u_c =
1+1/\sqrt{q}$ \cite{wu:82a}. Since $\langle b/N\rangle = p u$ (see, e.g.,
Ref.~\cite{weigel:02a}), one finds $\langle b/2N\rangle = 1/2$ for critical $0\le
q\le 4$. Here, $N$ denotes the total number of vertices. Thus, the bond occupation
probability of the critical RCM corresponds to the pure bond percolation threshold,
irrespective of $q$. For random bond selection, this results in constant and equal
fractions of insertions and deletions. As mentioned above in the context of the BFS
technique, the fractions of internal versus external edges are different from zero
for all values of $q$. Hence, it is the most expensive operation which dominates the
asymptotic scaling behavior, and we hence expect $\sim L^{\gamma/\nu}$ scaling for
the union-and-find implementation, although all insertion moves are performed in
constant time.

\subsubsection{Dynamic connectivity algorithm}

While union-and-find uses data structures that allow for insertions and connectivity
checks in constant time, there exist modified data structures for which also edge
deletions are supported without the need of an expensive $\sim L^{\gamma/\nu}$
rebuild operation in case of an external edge. A number of such ``fully dynamic''
graph algorithms has been discussed in the computer-science literature
\cite{henzinger:99,holm:01}. In the following, we refer to such approaches as {\em
  dynamic connectivity\/} (DC) algorithms. The advantage in run-time for the deletion
of edges is paid for in terms of increased efforts for edge insertion. We use the
approach suggested in Ref.~\cite{holm:01} which is deterministic and features
amortized runtimes of $\mathcal{O}(\log{N})$ for connectivity queries and
$\mathcal{O}(\log^2{N})$ for deletions and insertions on graphs of $N$ nodes. The
time complexity for connectivity queries depends on the underlying binary search tree
used to encode the graphs. In our case we used splay trees \cite{sleator:85} which
result in the amortized bound. An identical worst-case bound holds for balanced
binary search trees \cite{holm:01}.

The algorithm of Holm {\em et al.\/} \cite{holm:01} is based on a reduction of the
set of edges to be considered by focusing on a spanning forest $F(G)$ of a given
graph $G$, which encodes the same connectivity equivalence relation but has less
edges as no cycles occur \cite{gibbons:book}. This separates the set of edges into
``tree'' edges $e \in F(G)$ and ``non-tree'' edges $e \notin F(G)$. Given this
separation, one then stores the spanning trees of all components and augments them
with information about incident non-tree edges. This edge separation already reduces
the number of expensive operations, as any edge $e \notin F(G)$ must be an internal
edge, such that $\Delta n = 0$. The only potentially expensive cases remaining are
the deletion of $e \in F(G)$ and the insertion of an external edge into $F(G)$. For
the first case, one notes that the fact that the spanning tree is fragmented by the
removal of $e \in F(G)$ does not imply that also the cluster on the original graph is
split by this operation as there might be non-tree edges still connecting the
parts. The concept of tree- and non-tree edges is visualized in
Fig.~\ref{fig:edge_vis}.

In order to efficiently search for a replacement edge in the set of all non-tree
edges, another edge separation is introduced.  This is achieved by associating a
level in the range $0,1,\cdots,l_{\mathrm{max}} \equiv \lfloor\log N\rfloor$ to
every edge (both tree and non-tree).  Given the levels of all edges, one then
maintains a spanning forest $F_i$ for $i = 0,1,\cdots,l_{\mathrm{max}}$ of all edges
with $l(e) \ge i$, thus $F = F_{0} \supseteq F_1 \cdots\supseteq
F_{l_{\mathrm{max}}}$, cf.\ the illustration in Fig.~\ref{fig:edge_vis}. An important
part of the algorithm is that the levels are not fixed but they are partially changed
after every tree-edge deletion to ensure the following two properties \cite{holm:01}:
\begin{enumerate}
\item The maximum cluster size at level $i$ is $\lfloor N/2^i\rfloor$.  This implies
  $l_\mathrm{max} \equiv \lfloor\log N\rfloor$.
\item If an edge at level $i$ is deleted, then a possible replacement edge can only
  be found in levels $l \le i$.
\end{enumerate}
To efficiently save and manipulate spanning trees we represent each tree edge $(i,j)$
by two directed edges ($\equiv$ arcs) $(i \rightarrow j)$ and $(j \rightarrow i)$ and
every vertex $i$ by a loop $(i \rightarrow i)$.  Given all directed edges we
construct one of possibly many Euler tours of this directed graph, i.e., a cycle that
traverses every edge exactly once. We split the tour at one arbitrary point to save
it as a sequence of traversed arcs and loops \cite{tarjan:97}. The fact that we now
linearized the tree by mapping it to a list or sequence allows to efficiently store
it in a balanced binary search tree. Edge insertions and deletions then translate
into manipulations using cuts and links on Euler tours. By using a form of
self-adjusting binary search trees named ``splay trees'' \cite{sleator:85} we were
able to do all the operations on trees in amortized runtime of $\mathcal{O}(\log
N)$. Other types of trees might be used alternatively, see e.g.,
\cite{cormen:09,knuth:vol3,martinez:98}. Intuitively speaking, we hence have
$\mathcal{O}(\log N)$ levels of edges with runtime of $\mathcal{O}(\log N)$ per
level, resulting in the quoted $\mathcal{O}(\log^2 N)$ amortized runtime for
deletions and insertions. A detailed comparison and run-time analysis of different DC
algorithms can be found in Ref.~\cite{iyer:01}. For simulations of the RCM, we can
therefore perform each operation in a run-time asymptotically proportional to $\log^2
L$, hence clearly outperforming the other approaches at criticality, cf.\
Tab.~\ref{tab:scaling}.

\subsubsection{Behavior off criticality}

The run-time bounds summarized in Table \ref{tab:scaling} apply to simulations of the
RCM at criticality. In the high-temperature, non-percolating regime $p < p_c$ all
clusters are finite. Hence, the implementations based on BFS and UF allow to perform
insertions and deletions in asymptotically constant time there. For temperatures
below the transition, equivalent to $p > p_c$, the behavior is a bit more
complicated. In this case clusters become compact, so cluster masses scale
proportional to $L^2$. For SBFS, operations on external edges require complete
cluster traversal, leading to $L^2$ run-time scaling there. Through the dense nature
of clusters, a re-connecting path replacing an internal edge will, with probability
one, be only of finite length independent of $L$. Similarly, if an external edge
connects two distinct clusters, at least one of them will be small, i.e., of $L$
independent size. Hence, a constant run-time per operation is expected for IBFS and
$p>p_c$. In the UF implementation again complete cluster traversal is necessary in
some cases, giving an $L^2$ bound. The asymptotic scaling of DC equals $\log^2 L$,
independent of $p$, although the absolute run-times will of course depend on
temperature.

\section{Simulations and results\label{sec:results}}

To gauge the efficiency of the connectivity implementations and confirm the validity
of the asymptotic analysis presented above, we implemented Sweeny simulation codes
based on these different approaches and subjected them to a careful analysis of
run-times as a function of system size and $q$.

\subsection{Autocorrelation times and efficiency\label{sec:autocorrelations}}

Besides the run-time for individual bond operations discussed above in
Sec.~\ref{sec:algorithms}, the efficiency of the bond algorithm is ultimately
determined by the speed of decorrelation of the Markov chain, i.e., the
autocorrelation times. Consider the autocorrelation function of the measurements
$\mathcal{O}_s$ at time lag $t$,
\begin{equation}
  \Gamma_{\mathcal{O}}(t) = \langle\mathcal{O}_s \mathcal{O}_{s+t}\rangle -
  \langle\mathcal{O}\rangle^2
\end{equation}
which, in equilibrium, is expected to be independent of the initial time $s$ due to
stationarity.  From the theory of Markov chains, $\Gamma_{\mathcal{O}}(t)$ is
expected to decay exponentially for large time lags, $\Gamma_{\mathcal{O}}(t) \sim
e^{-\vert t\vert/\tau_{\mathrm{exp},\mathcal{O}}}$, and one defines the {\em
  exponential autocorrelation time\/} \cite{sokal:97}
\begin{equation}
  \tau_{\mathrm{exp},\mathcal{O}} = \limsup_{t\rightarrow \pm \infty}  \frac{\vert t
    \vert}{ -\log{\vert \rho_{\mathcal{O}}(t)}\vert },
\end{equation}
where $\rho_{\mathcal{O}}(t) = \Gamma_{\mathcal{O}}(t)/\Gamma_{\mathcal{O}}(0)$ is
the normalized autocorrelation function. Unless the considered observable is
``orthogonal'' to the slowest mode, one expects a result independent of ${\cal O}$,
i.e., $\tau_{\mathrm{exp}} \approx \tau_{\mathrm{exp},\mathcal{O}}$.

The efficiency of sampling, on the other hand, is determined by the {\em integrated
  autocorrelation time\/},
\begin{equation}
  \tau_{\mathrm{int},\mathcal{O}} = \frac{1}{2} + \sum_{t=1}^{M}
  \rho_{\mathcal{O}}(t),
  \label{eq:tau_int}
\end{equation}
where $M$ is the length of the time series. This is seen by considering the variance
of the average $\bar{\cal O} = (\sum_i {\cal O}_i)/M$ used as an estimator for
$\langle {\cal O}\rangle$,
\begin{equation}
  \sigma_{\bar{\mathcal{O}}}^2 \approx \frac{\sigma^2_{\cal O}}{M/2 \tau_{\mathrm{int},\mathcal{O}}}.
  \label{eq:variance}
\end{equation}
Comparing this to the case of uncorrelated measurements, where
$\sigma_{\bar{\mathcal{O}}}^2 = \sigma^2_{\cal O}/M$, it is seen that the effective
number of independent measurements is reduced by a factor of $1/2
\tau_{\mathrm{int},\mathcal{O}}$ by the presence of autocorrelations. In contrast to
$\tau_{\mathrm{exp}}$, the integrated autocorrelation time generically depends on the
observable considered.

Close to criticality, dynamical scaling implies autocorrelation times diverging
according to $\tau_{\mathrm{exp}} \sim \xi^{z_{\mathrm{exp}}}$ and
$\tau_{\mathrm{int},\mathcal{O}} \sim \xi^{z_{\mathrm{int},\mathcal{O}}}$. For finite
systems, $\xi$ eventually becomes limited by $L$, resulting in the scaling form

\begin{equation}
  \tau \sim L^z.
  \label{eq:tau_scaling}
\end{equation}
For a purely exponential decay, exponential and integrated autocorrelation times
coincide. More generally, however, one has $\tau_{\mathrm{int},{\cal O}} \le
\tau_\mathrm{exp}$ and hence $z_{\mathrm{int},\mathcal{O}} \le
z_{\mathrm{exp}}$. Cases of a true inequality have been observed \cite{madras:88a}.

\begin{figure}[t]
  \begin{center}
    \includegraphics[width=\columnwidth]{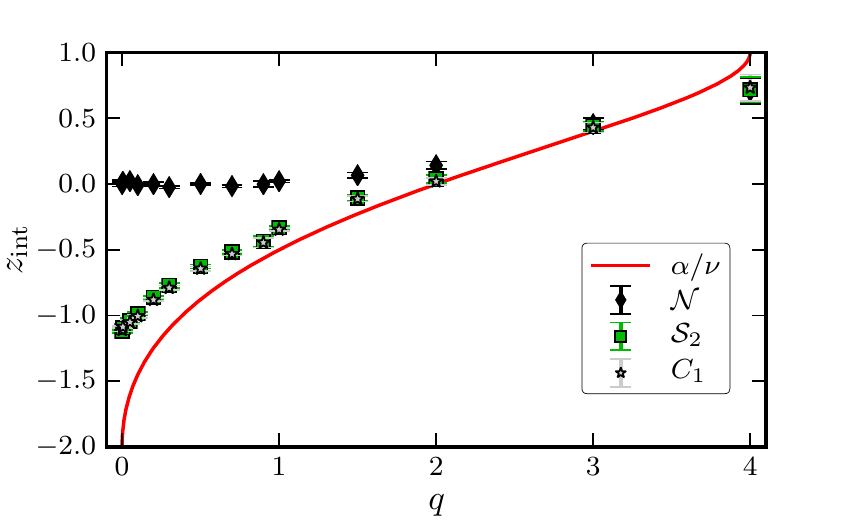}
    \caption{\label{fig:z_int_overall}(Color online) Estimates of the 2D dynamical critical exponents
      $z_{\mathrm{int},{\cal O}}$ for ${\cal O} = {\cal N}$, ${\cal S}_2$ and $C_1$,
      respectively, as a function of $q$. The line corresponds to the Li-Sokal bound
      $z_\mathrm{int} > \alpha/\nu$ \cite{li:89}.}
  \end{center}
\end{figure}

A number of different techniques for the practical estimation of autocorrelation
times have been discussed in the literature. We use a direct summation of the
autocorrelation function according to Eq.~\eqref{eq:tau_int}. Since, for any time
lag, the estimated autocorrelation function adds a constant amount of noise per term,
a cutoff $\Lambda$ needs to be introduced to ensure a finite variance of the
estimator for the autocorrelation time \cite{priestley:book}. A technique with an
adaptive summation window has originally been suggested in Ref.~\cite{madras:88a}. We
use a modification of this approach discussed more recently in
Ref.~\cite{wolff:04}. Determining $\Lambda$ involves a tradeoff between the bias for
small cutoffs and the exploding variance as $\Lambda\rightarrow M$. This balance is
struck here by numerically minimizing the quantity
\begin{equation}
\exp{\left( -\Lambda/\tau\right)} + 2 \sqrt{\Lambda/M},
\end{equation}
which is proportional to the sum of the relative statistical and systematic
errors. We find this procedure to yield stable results throughout. In particular, the
estimates of $\tau_\mathrm{int}$ are consistent with those found from an alternative
jackknifing technique \cite{weigel:10,efron:book}.

To achieve an appropriate judgment of Sweeny's algorithm in the different
implementations against other approaches of simulating the RCM, we need to combine
the information contained in the autocorrelation times with those of the run-times
for the different operations involved. We hence compare the effective run-time to
create a statistically independent sample of a given observable, where independence
is understood in the sense of Eq.~\eqref{eq:variance}. To this end, we consider the
effective runtime per edge,
\begin{equation}
  T_{\mathcal{O}} \equiv \tau_{\mathrm{int},\mathcal{O}} \bar{t},
\label{eq:eff_statTime}
\end{equation}
where $\tau_{\mathrm{int},\mathcal{O}}$ is measured in sweeps and $\bar{t}$  is the
runtime per edge operation, averaged over a sufficiently long
simulation. Obviously, this quantity is hardware specific, yet by looking at the ratio 
of two different implementations we expect the specific hardware-dependence to be
small \footnote{Caching effects might lead to unexpected behavior in some range of
  system sizes, but asymptotically these effects should be irrelevant.}.
 
\subsection{Observables and implementation}

\begin{figure}[t]
  \begin{center}
    \includegraphics[width=\columnwidth]{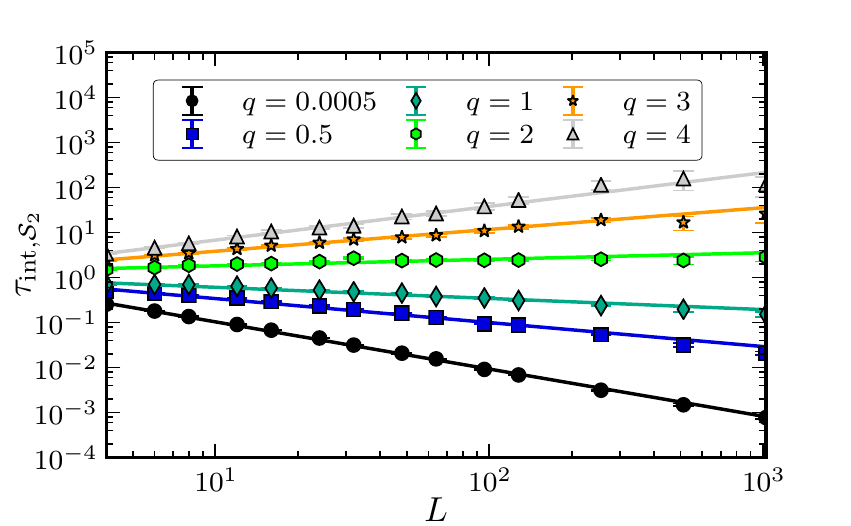}
    \caption{\label{fig:tau_int}(Color online) Integrated autocorrelation times as a function of
      system size for the susceptibility $\mathcal{S}_2$ for $q$ ranging from
      $0.0005$ up to $q=4$. The lines show fits of the functional form
      \eqref{eq:tau_scaling} to the data. }
  \end{center}
\end{figure}

As $\tau_{\mathrm{int},\mathcal{O}}$ depends on $\mathcal{O}$, any conclusions about
the relative efficiency of the Sweeny and Swendsen-Wang-Chayes-Machta dynamics for
the RCM might depend on the observable under consideration. It is therefore important
to study a number of different quantities covering the energetic and magnetic sectors
of the model. For the energetic sector, we studied the number ${\cal N} \equiv b$ of
active bonds. In the magnetic sector, we considered the sum of squares of cluster
sizes $\mathcal{S}_2 \equiv \sum_{i=1}^n \vert {\cal C}_i \vert^2$, and the size of
the largest component $C_{1} \equiv \max_{i=1}^n \vert {\cal C}_i \vert$, where
${\cal C}_i$ denotes the $i$th cluster resulting from the set of active bonds.  These
observables of the RCM are related to standard observables of the Potts
model~\cite{weigel:02a,garoni:11}, namely the internal energy per spin $u =
\langle\mathcal{N}\rangle/(L^d p)$, the susceptibility $\chi =
\langle\mathcal{S}_2\rangle/L^d$, and the order parameter $m = \langle
C_1\rangle/L^d$.

We implemented a code for the Sweeny update using the four variants of connectivity
algorithms discussed above in Sec.~\ref{sec:algorithms}, i.e., sequential
breadth-first search (SBFS), interleaved BFS (IBFS), union-and-find (UF), as well as
dynamic connectivities (DC). Our program was implemented in C and compiled with the
GNU Compiler Collection (GCC) 4.5.0 at -O2 optimization level on an Intel Xeon E4530
$2.66$ GHz.  The available memory was $16$ GB. These specifications resulted in speed
ups through caching effects -- mainly for the DC implementation, see the discussion
below -- for system sizes $L \le 48$, which approximately corresponds to a memory
footprint of the size of the L2 cache of $6$ MB for the DC implementation.

The implementations based on BFS and union-and-find both require memory scaling as
$\mathcal{O}(L^2)$. The DC code, on the other hand, requires to maintain
$\mathcal{O}(\log L)$ separate forests, leading to a total memory requirement of
$\mathcal{O}(L^2\log L )$ \cite{iyer:01}. In practise for a system of size $L=1024$
and $q=2$ at criticality, the BFS and UF implementations required around $10$--$15$
MByte, whereas the DC code used approximately $3.5$ GByte.

The random number stream was generated by a GSL implementation \footnote{GNU
  Scientific Library, http://www.gnu.org/software/gsl} of the Mersenne twister or
MT19937 generator \cite{matsumoto:98} which has a very large period of $2^{19937}
\approx 10^{6000}$ and, more importantly, was shown to be equi-distributed in less
then $623$ dimensions.

\begin{figure}[t]
  \begin{center}
    \includegraphics[width=\columnwidth]{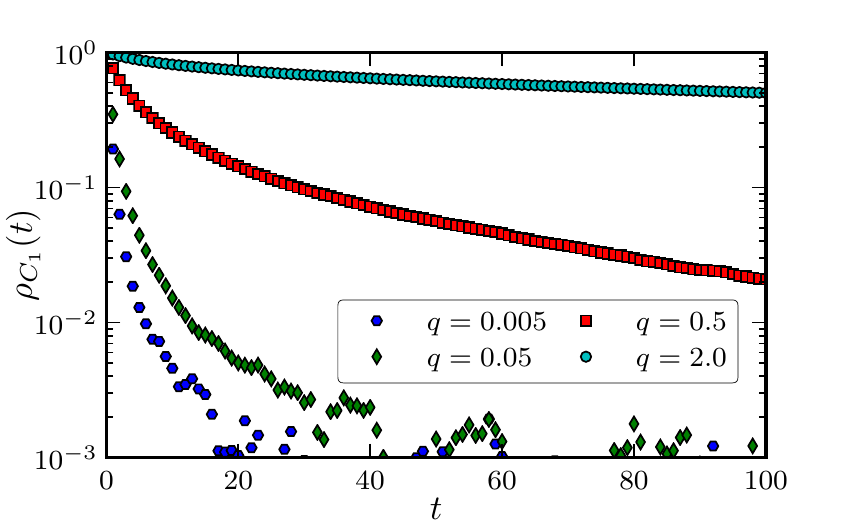}
    \caption{\label{fig:rhos}(Color online) Normalized critical autocorrelation
      function for the order parameter $C_1$ and several values of $q$. The linear
      system size is $L=64$. Time is counted in units of $L$ single bond moves here,
      such that $L = 64$ time steps correspond to a lattice sweep.  }
  \end{center}
\end{figure}

\subsection{Simulation results}

In order to test the predicted asymptotic run-time behavior and determine the dynamic
critical behavior of Sweeny's algorithm, we performed a series of simulations of the
RCM on the square lattice. To cover the complete range of systems with continuous
phase transitions, we chose $q=0.0005$, $0.005$, $0.05$, $0.1$, $0.2$, $0.3$, $0.5$,
$0.7$, $0.9$, $1$, $1.5$, $2$, $3$ and $4$. Simulations were performed for a series
of $14$ system sizes ranging from $L = 4$ up to $L = 1024$. All simulations were done
at the exact critical coupling $v_c = \sqrt{q}$ \cite{wu:82a}. While, typically,
observable measurements in Markov chain Monte Carlo are taken after every {\em
  lattice sweep\/} of updates \cite{binder:book2}, for the case of the bond algorithm
and relatively small values of $q$, it turned out that the fast decorrelation leads
to autocorrelation times way below a single sweep. To resolve these effects, we hence
changed the measurement interval to multiples of $L$, which turned out to be a good
compromise between the visibility of correlations and the resulting lengths of time
series. This setup resulted in at least $2^{20}$ and up to $2^{23}$ measurements for
some system sizes. As a result, our simulations were at least $10^3$, and at best
$10^5$ times the relevant time scale $\tau_\mathrm{exp}$.  Consistent with Sweeny's
original observation, we were able to equilibrate all runs in less then $200$ sweeps.

\subsubsection{Dynamical critical behavior}

We first considered the behavior of the energy-like observable ${\cal N}$,
determining (integrated) autocorrelation times according to Ref.~\cite{wolff:04} with
resulting cutoff parameter $\Lambda_{\cal N}$ (see the discussion in
Sec.~\ref{sec:autocorrelations} above). In order to extract the corresponding
dynamical critical exponents we fitted a power-law of the form $\tau_\mathrm{int}
\sim L^{z_\mathrm{int}}$ to the data, omitting some of the smallest system sizes to
account for scaling corrections. The quoted errors on fit parameters correspond to an
interval of one standard deviation. The resulting estimates of $z_{\mathrm{int},{\cal
    N}}$ are shown in Fig.~\ref{fig:z_int_overall} and the numerical values are
summarized in Table \ref{tab:z_ints}. Note that $z_{\mathrm{int},{\cal N}} \approx 0$
for $q\le 2$. This is in agreement with a key result due to Li and Sokal, providing a
lower bound for the autocorrelation times of $\mathcal{N}$ and its corresponding
dynamical exponents \cite{li:89,garoni:11}
\begin{equation}
\tau_{\mathrm{exp},\mathcal{N}} \gtrsim \tau_{\mathrm{int},\mathcal{N}} \ge C_v \;\Rightarrow\;
z_{\mathrm{exp},\mathcal{N}} \ge z_{\mathrm{int},\mathcal{N}}\ge  \alpha/\nu,
\label{eq:lisokal}
\end{equation}
where $C_v$ is the specific heat and $\alpha/\nu$ the associated finite-size scaling
exponent. While this result was originally derived for the Swendsen-Wang dynamics, it
was also shown to hold for the Sweeny algorithm \cite{li:89,deng:07}. As $\alpha/\nu
\le 0$ for $q\le 2$ \cite{wu:82a}, our data are consistent with this bound and
indicate that it is close to being tight for the Sweeny dynamics on the square
lattice for $q\ge 2$, cf.\ Fig.~\ref{fig:z_int_overall}. The values for $q=4$ appear
to violate this bound, but we attribute these deviations to the logarithmic
corrections expected for this particular value of $q$, preventing us from seeing the
truly asymptotic behavior in the regime of system sizes considered here.

We next turned to the magnetic observables ${\cal S}_2$ and $C_1$. Deng {\em et
  al.\/} \cite{deng:07} first showed that under the Sweeny dynamics the
susceptibility ${\cal S}_2$ exhibits a surprisingly fast decorrelation on short
timescales and, in particular, the corresponding integrated autocorrelation time
$\tau_{\mathrm{int},\mathcal{S}_2}$ {\em decreases\/} with increasing system size,
indicating a negative value of the corresponding dynamical critical exponent
$z_{\mathrm{int},\mathcal{S}_2}$. This phenomenon of {\em critical speeding up\/} is
also clearly seen in our data as is illustrated in the plot of the $L$ dependence of
the autocorrelation times of ${\cal S}_2$ in Fig.~\ref{fig:tau_int}. A rather similar
behavior is found for the order parameter $C_1$. The initial fast decay of
correlations is illustrated in Fig.~\ref{fig:rhos}, where it is clearly seen that
$C_1$ is completely decorrelated in less than a single sweep for $q\lesssim 2$. Note
that the measurements along the Markov chain are still correlated, but with a system
size scaling weaker than $L^2$ such that the impression of a complete decorrelation
appears on the scale of sweeps.

\begin{figure}[t]
  \begin{center}
    \includegraphics[width=\columnwidth]{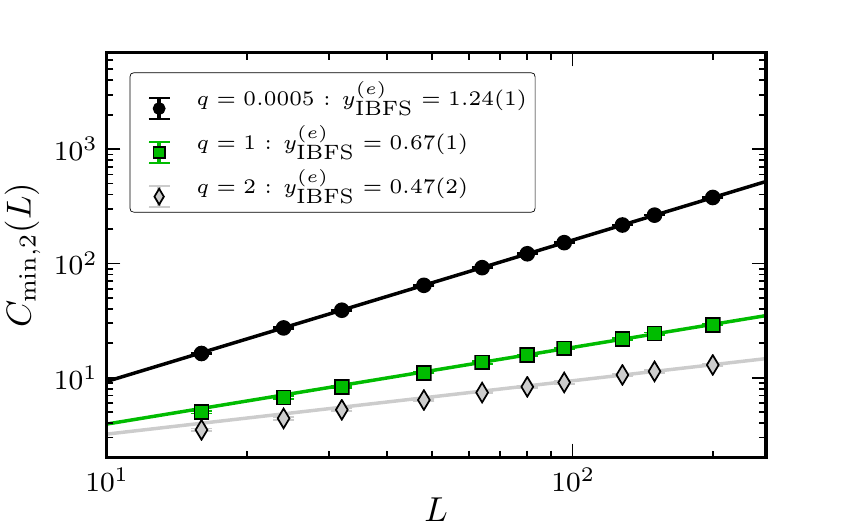}
    \caption{\label{fig:smaller_size}
        (Color online) Number of vertices $C_{\mathrm{min},2}$ of the smaller of two adjacent
      clusters as visited in a 2D simulation using IBFS following (preceding) the
      deletion (insertion) of an external edge. The lines show fits of the
      power-law form $C_{\mathrm{min},2} = A L^{y_{\textrm{IBFS}}^{(e)}}$ to the
      data.}
  \end{center}
\end{figure}

The values of $z_{\mathrm{int},\mathcal{S}_2}$ resulting from power-law fits to the
autocorrelation times are compiled in Table \ref{tab:z_ints}. As is clearly seen from
the plot of the data in Fig.~\ref{fig:tau_int}, we find
$z_{\mathrm{int},\mathcal{S}_2} \le 0$ for $q\le 2$. Regarding the dynamical critical
exponents in the magnetic sector, we find $z_{\mathrm{int},C_1} =
z_{\mathrm{int},\mathcal{S}_2}$ within our error bars. The authors of
Ref.~\cite{deng:07} suggested to determine $z_{\mathrm{int},\mathcal{S}_2}$ from a
data collapsing procedure using a two-time scaling ansatz combining the fast initial
decay with a slower exponential mode for longer times. In our studies, however, we
found this approach to yield rather unstable results and thus chose, instead, to
perform more conventional fits to a power law.

We note that, in line with Ref.~\cite{garoni:11} we used the summation cutoff
$\Lambda_{\mathcal{N}}$ of the observable $\mathcal{N}$ for all observables, because
the magnetic observables ${\cal S}_2$ and $C_1$ with their fast initial decay would
lead to very small, sub-asymptotic cutoffs if the rule of Ref.~\cite{wolff:04} would
be directly applied. We also checked that the estimators for
$\tau_{\mathrm{int},\mathcal{S}_2}$ and $\tau_{\mathrm{int},C_1}$ were on a plateau
so that a change in the summation window mainly influences the variance of the
estimator, which is monotonically increasing with $\Lambda$.

\begin{table*}[tb!]
  \caption{\label{tab:exp_runtime}
    Run-time scaling exponents in two dimensions according to Eq.~\eqref{eq:runtime_scaling} for the
    SBFS, IBFS and the UF implementation. The scaling exponents $\gamma/\nu$ and $d_F-x_2$ are shown for reference
    and comparison. The $y$ exponents correspond to scaling $\sim L^{y}$ of the
    number of vertices touched  in a
    sequential (SBFS) and interleaved (IBFS) breadth-first cluster traversal for internal (i)
    and external (e) edges, respectively.
  }
  \begin{ruledtabular}
    \begin{tabular}{cccccccccc}
        $q$ & $\kappa_\mathrm{SBFS}$&$\kappa_\mathrm{IBFS}$ & $\kappa_\mathrm{UF}$ &
        $y_{\textrm{IBFS}}^{(i)}$ & $y_{\textrm{IBFS}}^{(e)}$&
        $y_{\textrm{SBFS}}^{(i)}$& $y_{\textrm{SBFS}}^{(e)}$& $d_F - x_2$ &$\gamma/\nu$ \\ \hline \hline

        0.0005&    $1.81(1)$& $1.18(3)$&$1.54(6)$   &$1.25(1)  $&$1.24(1)$&$1.25(1)$&$1.99(1)$&	1.23407	&	1.99296	\\
        0.005&     $1.80(2)$& $1.14(2)$&$1.70(6)$   &$1.22(1)  $&$1.21(1)$&$1.22(1)$&$1.98(1)$&	1.20021	&	1.97823	\\
        0.05&      $1.74(1)$& $1.02(3)$&$1.77(4)$   &$1.10(1)  $&$1.11(1)$&$1.11(1)$&$1.93(1)$&	1.09783	&	1.93580	\\
        0.1&       $1.72(1)$& $0.97(2)$&$1.78(3)$   &$1.05(1)  $&$1.05(1)$&$1.06(1)$&$1.91(1)$&	1.03881	&	1.91284	\\
        0.5&       $1.65(2)$& $0.71(3)$&$1.77(2) $  &$0.82(1)  $&$0.82(1)$&$0.82(1)$&$1.83(1)$&	0.80768	&	1.83449	\\
        0.7&       $1.63(2)$& $0.69(4)$&$1.77(4) $  &$0.75(2)  $&$0.76(1)$&$0.76(2)$&$1.81(1)$&	0.73541	&	1.81407	\\
        1&         $0$& $0$      &$1.74(2)  $       &$0.66(1)  $&$0.67(1)$&$0.67(1)$&$1.79(1)$&	0.64583	&	1.79167	\\
        1.5&       $1.56(2)$& $0.43(2)$&$1.71(3)$   &$0.55(2)  $&$0.56(1)$&$0.57(2)$&$1.75(1)$&	0.52298	&	1.76644	\\
        2&         $1.57(2)$& $0.35(3)$&$1.68(3)$   &$0.46(2)  $&$0.47(2)$&$0.48(3)$&$1.73(1)$&	0.41667	&	1.75000	\\
        3&         $1.52(4)$& $0.19(1)$&$1.67(2)$   &$0.32(3)  $&$0.30(4)$&$0.35(3)$&$1.69(2)$&	0.21667	&	1.73333	\\
        4&         $1.42(4)$& $0.13(2)$&$1.64(7)$   &$0.22(11) $&$0.23(1)$&$0.26(1)$&$1.68(1)$&	-0.12500	&	1.75000		
    \end{tabular}
  \end{ruledtabular}
\end{table*}

Comparing our results for the Sweeny dynamics to the Swendsen-Wang algorithm, we note
that, apart from the slightly smaller dynamical critical exponent for the former, we
also find somewhat smaller amplitudes in the $\tau_\mathrm{int} =
AL^{z_\mathrm{int}}$ scaling for the bond algorithm. Hence for $L=256$ we find, e.g.,
$\tau_{\mathrm{int},{\cal S}_2} \approx 0.1$ ($q=1$) and $\approx 10$ ($q=3$) for the
Sweeny update, while values of $\tau_{\mathrm{int},{\cal S}_2} \approx 0.5$ ($q=1$)
and $\approx 36$ ($q=3$) are found for the Swendsen-Wang update.

\subsubsection{Run-time scaling}

As discussed above, the relevant time scales for a comparison of the bond algorithm
against other approaches depend on both, the statistical decorrelation as well as the
run-time scaling of the elementary operations. We therefore analyzed average
run-times for bond updates in the Sweeny algorithm with the different implementations
of connectivity updates discussed in Sec.~\ref{sec:connectivity}. 

\begin{figure}[t]
  \begin{center}
    \includegraphics[width=0.5\textwidth]{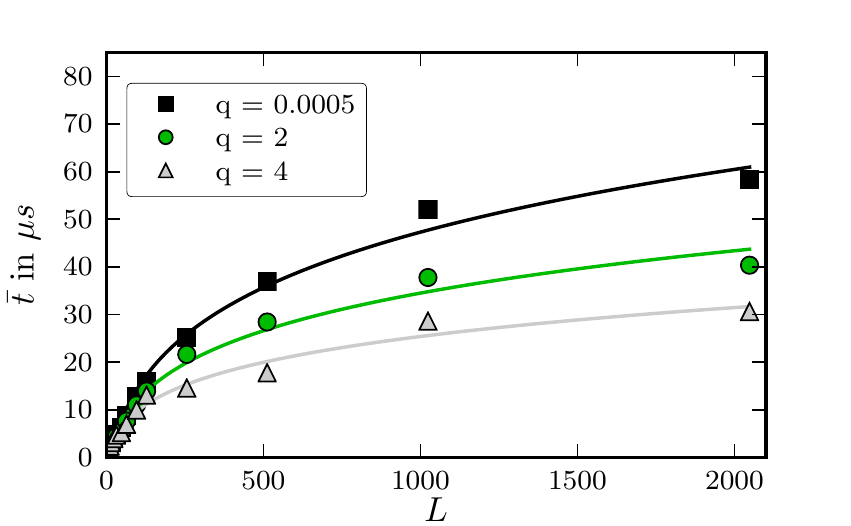}
    \caption{\label{fig:runtime_sy_sy}(Color online) Average run-time $\bar{t}$ for several values
      of $q$ and $L$ for the Sweeny update using a dynamic connectivity (DC)
      algorithm based on splay trees. The lines correspond to least-squares fits of
      the model \eqref{eq:DC_scaling} to the data.}
  \end{center}
\end{figure}

For the techniques based on BFS, we studied the number of steps required to complete
a connectivity check for the case of operations on internal and external edges,
respectively, for the SBFS and SBFS implementations. For internal edges, this
corresponds to the number of vertices touched by the BFSs until a re-connecting path
is found. For external edges such a path is not found and the search hence terminates
after a number of steps corresponding to the mass of either the first cluster (SBFS)
or the smaller cluster (IBFS).  Checking the number of steps for operations on
internal and external edges for SBFS and IBFS, respectively, we used power-law fits
according to $\sim L^y$ to extract estimates of the four exponents
$y_{\textrm{SBFS}}^{(i)}$, $y_{\textrm{SBFS}}^{(e)}$, $y_{\textrm{IBFS}}^{(i)}$, and
$y_{\textrm{IBFS}}^{(e)}$. The fit results are collected in Table
\ref{tab:exp_runtime}. The data and corresponding fits for the case of the number of
steps $C_{2,\mathrm{min}}$ relevant for the operation on an external edge with IBFS
are shown in Fig.~\ref{fig:smaller_size}. The exponents $y$ follow the asymptotic
values $\gamma/\nu$ and $d_F-x_2$, respectively, derived above in Sec.~\ref{sec:bfs}
and also listed in Table \ref{tab:exp_runtime} for comparison.

For the total average run-time per bond operation, we asymptotically expect power-law
behavior as well,
\begin{equation}
  \label{eq:runtime_scaling}
  \bar{t} \sim L^\kappa.
\end{equation}
This assumption in general describes well our data --- with only minor deviations for
smaller system sizes due to caching effects. For SBFS, we expect the different
asymptotic scaling behavior for operations on internal and external edges,
respectively, to result in an effective run-time exponent $\kappa$ somewhere in
between the exponents $d_F-x_2$ and $\gamma/\nu$ relevant to operations on internal
and external edges, respectively (recall that internal and external edges occur in
constant fractions). Our estimates of $\kappa_\mathrm{SBFS}$ listed in Table
\ref{tab:exp_runtime} are in line with these expectations. We have no doubt, however,
that the asymptotically expected $\kappa_\mathrm{SBFS} = \gamma/\nu$ ultimately holds
for sufficiently large systems. For the interleaved case, on the other hand, all four
operation types exhibit $y = d_F-x_2$, and we hence find $\kappa_\mathrm{IBFS}$
consistent with $d_F-x_2$ already for the system sizes considered here, cf.\ Table
\ref{tab:exp_runtime}.

The analysis of the run-time behavior for the union-and-find approach is more subtle
as the insertion of edges is performed in constant time, whereas the deletion of
edges incurs an effort proportional to $L^{d_F-x_2}$ and $L^{\gamma/\nu}$ for
internal and external edges, respectively, cf.\ Table \ref{tab:scaling}. As a
consequence of the different scaling of individual operations, the effective run-time
scaling exponent $\kappa_\mathrm{UF}$ according to Eq.~\eqref{eq:runtime_scaling} is
again found to be smaller than the expected limiting value $\gamma/\nu$, see the
values compiled in Table \ref{tab:exp_runtime}.



The scaling of run-times per step for our implementation of the dynamic connectivity
algorithm and a representative selection of $q$-values is shown in
Fig.~\ref{fig:runtime_sy_sy}. We find a sub-algebraic growth and, according to the
asymptotic run-time bounds derived in Ref.~\cite{holm:01}, we fitted the functional
form
\begin{equation}
  \label{eq:DC_scaling}
  \bar{t}(L) =  a\log^2 L + b \log L + c
\end{equation}
to the data. The fits resulted in $c\approx 0$ such that we fixed $c=0$ in the
following. Somewhat surprisingly, our fits yield $b < 0$; we interpret this as a
result of the presence of correction terms and the amortized nature of the run-time
bounds leading to the asymptotic scaling only being visible for very large system
sizes. Similar observations have been reported for general sets of inputs in
Ref.~\cite{iyer:01}. Considering the ratio $a/b$, we find that its modulus increases
with $q$, yielding a value of $\approx 0.3$ for $q=0.0005$ and $\approx 0.71$ for
$q=2$. This corresponds to the increasing fraction of non-tree edges for increasing
$q$, resulting in an increase of traversals of the edge level hierarchy with the
associated ${\cal O}(\log^2 L)$ complexity. Irrespective of that, as a consequence of
the larger number of cluster-splitting operations the total run-time is found to be
largest for small $q$, cf.\ Fig.~\ref{fig:runtime_sy_sy}.

\begin{figure*}
  \begin{tabular}{@{}lll@{}}
    \includegraphics[width=\columnwidth]{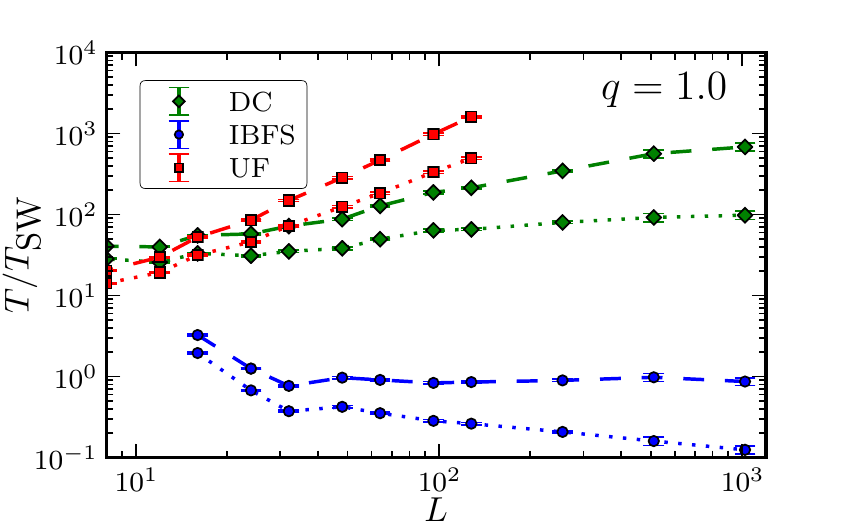} &\hspace{0.05\textwidth} &
    \includegraphics[width=\columnwidth]{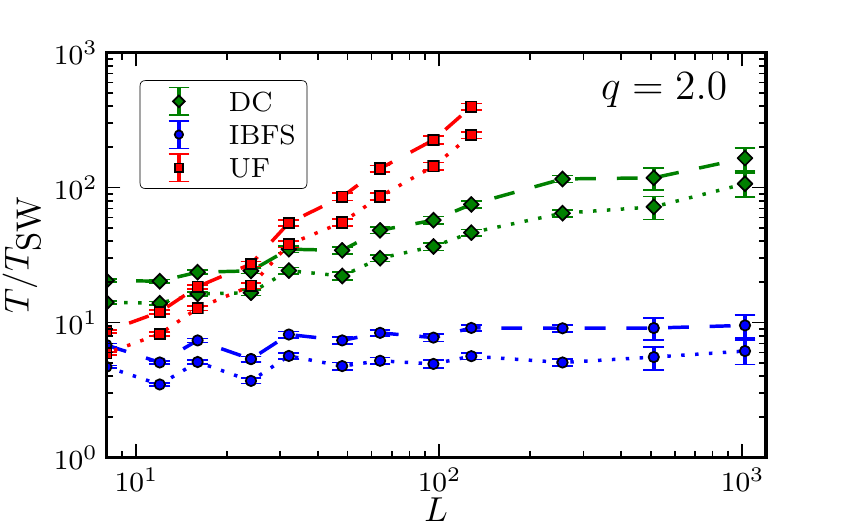}   \\ \vspace{0.01\textwidth} \\
    \includegraphics[width=\columnwidth]{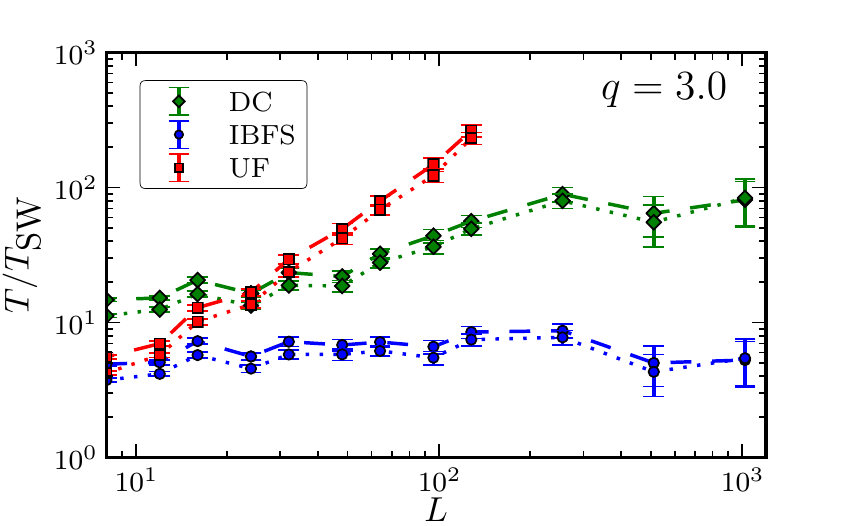} &\hspace{0.05\textwidth} &
    \includegraphics[width=\columnwidth]{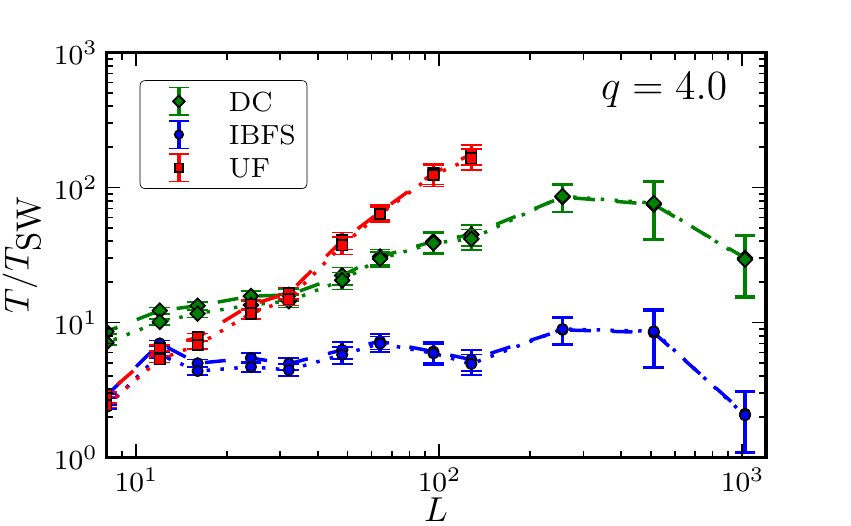}   \\
  \end{tabular}
  \caption{\label{fig:effective_runtimes}(Color online) Effective run-times $T$ according to
    Eq.~\eqref{eq:eff_statTime} for the different implementations relative to the
    time $T_\mathrm{SW}$ of the Swendsen-Wang algorithm \cite{swendsen-wang:87a}. Dashed
    lines correspond to the run-time to generate an independent sample of the
    observable $\mathcal{N}$ and dotted lines to samples of $\mathcal{S}_2$.}
\end{figure*}

We also investigated the effect of the unconditional acceptance of proposed updates
for the Metropolis rule as discussed in Sec.~\ref{sec:sweeny} above. This adds
another $q$ dependent, but system-size independent, element to the run-time
scaling. Such unconditional moves can save significant computational effort in case
no data structures have to be updated after move acceptance. This is the case for the
algorithms based on BFS which are ``stateless'' in the sense that no explicit record
of connectivity is kept. Unconditional insertion or removal of edges in this case
does not entail any further computational effort. On the contrary, unconditional
insertion or removal lead to data-structure updates for the union-and-find and DC
implementations. As a consequence, we find a constant speed-up for the BFS based
implementations proportional to $1/[1-\min(\sqrt{q},1/\sqrt{q})]$. In the singular
case $q=1$, BFS performs all edge updates in constant time as all insertions and
deletions can be performed unconditionally such that no cluster traversals are
necessary. On the contrary, no performance improvement from unconditional moves is
observed for the more elaborate UF and DC implementations.

We note that, for all implementations, the average run-time per bond operation
depends quite strongly on $q$. This is, on the one hand, due to the $q$ dependence of
the fraction $r_\mathrm{ext}$ of external edges reaching from $r_\mathrm{ext} = 1$
for $q\to 0$ down to $r_\mathrm{ext} = 0.33$ for $q = 4$. For the case of the
BFS implementations, an additional $q$ dependence is introduced through the
unconditional moves as discussed above.

\subsubsection{Overall efficiency}

As discussed above in Sec.~\ref{sec:autocorrelations}, the relevant measure for the
overall efficiency of various implementations of cluster algorithms is the total
run-time for the generation of a statistically independent sample according to
Eq.~\eqref{eq:eff_statTime}. We compared the effective run-times of all three
implementations of the bond algorithm with a reference code for the Swendsen-Wang
dynamics. As the dynamical critical exponents $z_\mathrm{int}$ for the Sweeny update
are found to be smaller than those of the Swendsen-Wang-Chayes-Machta dynamics, see
Refs.~\cite{deng:07,deng:07a,garoni:11} and Table \ref{tab:z_ints}, we can expect
asymptotically more efficient simulations for cases where the run-time exponent
$\kappa < z_{\mathrm{int},{\cal O}}^\mathrm{SW}-z_{\mathrm{int},{\cal O}}$. Since any
poly-logarithm is dominated by $L^{\epsilon}$ with $\epsilon > 0$, this is clearly
the case, asymptotically, for the DC algorithm. From the data for
$z_{\mathrm{int},{\cal O}}^\mathrm{SW}-z_{\mathrm{int},{\cal O}}$ in Table
\ref{tab:z_ints} and those for $\kappa$ in Table \ref{tab:exp_runtime}, it is seen
that for $1 < q\le 4$, this condition is not met for the implementation based on
union-and-find. For the technique based on (interleaved) breadth-first-search, on the
other hand, such a case arises (for integer values $q$) only for $q=4$, where
$z_{\mathrm{int},{\cal O}}^\mathrm{SW}-z_{\mathrm{int},{\cal O}} \approx 0.16$ and
$\kappa_\mathrm{IBFS} = 0.13(2)$. These observations are corroborated by the plots of
the relative efficiencies shown in Fig.~\ref{fig:effective_runtimes}. For comparison,
we here show the results for the two observables ${\cal N}$ and ${\cal S}_2$ with
significantly different behavior for $q\le 2$. From the plots for the integer values
$q=2$, $3$ and $4$, it is clear that in absolute run-times IBFS is most efficient for
the range of system sizes $L \le 1024$ considered here. Hence, the asymptotic
advantage of the DC algorithm only shows for system sizes beyond this range. The
downturn of the ratio $T/T_\mathrm{SW}$ for the largest system sizes and $q=4$
observed for the BFS and DC codes might be an indication of the asymptotic run-time
advantage of Sweeny's algorithm over Swendsen-Wang with these connectivity algorithms
as discussed above. The results for the percolation case $q=1$ using IBFS, on the
other hand, are of exceptional nature as there the cost of bond operations is
completely independent of system size due to the effect of unconditional
acceptance. For the case of ${\cal S}_2$, one even finds a decrease of the relative
cost of generating a statistically independent sample as this observable profits from
the initial fast decorrelation or critical speeding up.

\begin{figure}[t]
  \begin{center}
    \includegraphics[width=0.5\textwidth]{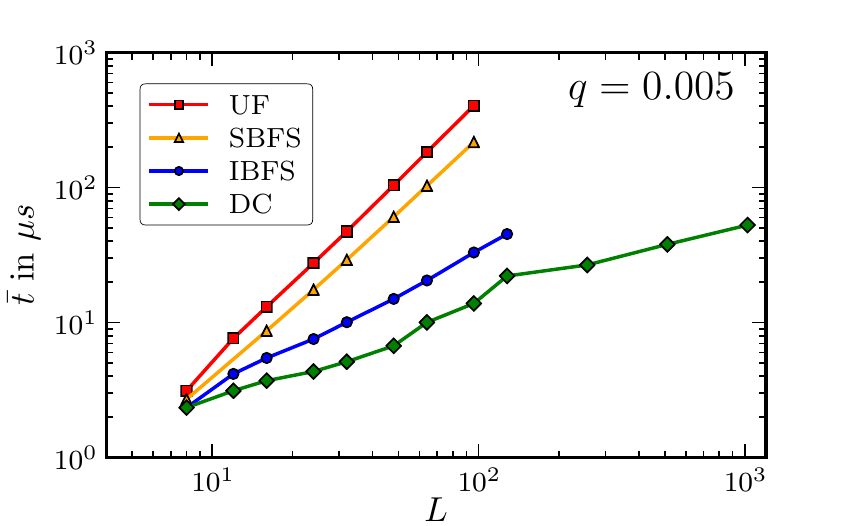}
    \caption{\label{fig:runtimes_impl_sweeny}
        (Color online) Run-time per edge operation of
      simulations of the $q=0.005$ square-lattice RCM 
      and the bond algorithm employing the SBFS, IBFS, UF and
      DC connectivity implementations, respectively.
    }
  \end{center}
\end{figure}

For the most relevant case $q\le 1$, where Sweeny's algorithm provides the only means
of simulation, we cannot compare to another algorithm. Instead, we present in
Fig.~\ref{fig:runtimes_impl_sweeny} a comparison of run times for the SBFS, IBFS, UF
and DC implementations for $q=0.005$. As here $\kappa_\mathrm{IBFS}$ is relatively
unfavorable, we observe a clear advantage for the DC algorithm which is significantly
more efficient than the other options in the full range of studied system sizes $4\le
L\le 1024$. UF is found to be even less efficient than SBFS here which might be
considered surprising in view of the fact that all insertions are performed at
constant cost and deletions have the same asymptotic run-time bounds as SBFS, see
Table \ref{tab:scaling}. This is easily understood, however, noting that the factor
of two gained for UF from the 50\% of operations performed in constant time is spent
again in having to traverse {\em both\/} clusters fully in case of external edge
deletions. Taking into account overheads for data-structure updates for UF, this
explains the slight disadvantage of UF over SBFS seen in
Fig.~\ref{fig:runtimes_impl_sweeny}.

\section{Conclusions\label{sec:concl}}

We have shown that it is possible to implement Sweeny's algorithm efficiently and in
a lattice and dimensionality independent way, using a dynamic connectivity (DC)
algorithm, in the sense that the runtime dependence on the system size is
poly-logarithmic and only contributes a correction to the statistical
$L^{z_\mathrm{int}}$ contribution of the runtime to create an effectively
uncorrelated sample. Compared to the Swendsen-Wang-Chayes-Machta algorithm, we also
find somewhat smaller dynamical critical exponents, leading to an overall
asymptotically more efficient simulation of the random-cluster or Potts model with
Sweeny's approach. In addition, the bond algorithm is the only known approach for
simulations in the regime $q < 1$ including, for instance, interesting $q\rightarrow
0$ limits such as the maximally connected spanning sub-graphs of
Ref.~\cite{jacobsen:04}.

We analyze in detail four implementations based on (sequential and interleaved)
breadth-first searches, on union-and-find data structures, and on the fully dynamic
connectivity algorithm suggested in Ref.~\cite{holm:01}, respectively. For each
implementation, we derive average run-time bounds for insertions and deletions of
internal and external edges, respectively, and deduce the overall asymptotic run-time
behavior. It is found that interleaved breadth-first searches, although relatively
unfavorable as compared to union-and-find and dynamic connectivities at first sight,
perform rather well due to the lack of an underlying data structure encoding the
connectivity of the clusters, in particular if connectivity queries are omitted
whenever possible due to accepting moves for which the drawn random number indicates
acceptance irrespective of the result of the connectivity query. The union-and-find
based implementation, on the other hand, although superior in asymptotic run-time in
three out of four cases of insertions or deletions of internal or external edges, shows
ultimately inferior performance due to the run-time scaling for deletions of external
edges that require full traversals of the involved clusters. The dynamic connectivity
algorithm, while asymptotically most efficient with a poly-logarithmic scaling of
run-times per operation, has rather large constants leading to somewhat weaker
performance than breadth-first search for the considered lattice sizes $L\le 1024$
and $q\ge 1$. For $q\ll 1$, on the other hand, where run-times are dominated by
operations on external edges, it outperforms the other implementations already for
small systems. We see significant room for further run-time improvements for the
dynamic connectivity algorithm, however, for instance by optimizations of the
underlying tree data structure or the implementation of additional heuristics as
indicated in the comparative study \cite{iyer:01}.  We note that due to the lack
of explicit connectivity information for the
breadth-first search approach it becomes more expensive than for the other techniques
to perform measurements of quantities such as cluster numbers or correlation
functions depending on the connectivity. As measurements are typically taken at most
once per sweep, however, any cost of at most $L^2$ operations for measurements
results in only ${\cal O}(1)$ amortized effort per bond operation.

The observed fast initial decorrelation for $q\le 2$ and quantities such as ${\cal
  S}_2$ and $C_1$ depending on cluster connectivity as illustrated in
Figs. \ref{fig:z_int_overall}, \ref{fig:tau_int} and \ref{fig:rhos} suggests that
there is an additional dynamical mechanism at play for such observables. As argued in
Ref.~\cite{deng:07}, this is due to a larger number of operations on external edges
for smaller values of $q$ which can lead to a large-scale change in the connectivity
structure through a single bond operation. The concentration of external edges,
bridges or fragmenting bonds drastically increases as $q$ is decreased from $4$ down
to the tree limit $q \to 0$ \cite{elci:prep} which is also clearly expressed in a
corresponding increase in the fractal dimension of ``red'' bonds
\cite{stanley:77}. It is currently not clear, however, why this effect only leads to
a change in dynamical critical behavior for $q\le 2$.

While we have restricted our attention to simulations on the square lattice, all
implementations discussed here are essentially independent of the underlying graph or
lattice, requiring only minimal adaptations for different situations. This makes our
approach significantly more general than the implementation originally suggested by
Sweeny \cite{sweeny:83} which is based on tracing loops on the medial lattice in two
dimensions. Additionally, the latter technique in two dimensions still suffers from
polynomial scaling of the run-time per edge operation \cite{deng:10}, such that a
poly-logarithmic implementation is asymptotically faster.

Until now we have only used the DC implementation for canonical simulations but as
proposed in Ref.~\cite{weigel:10d} an interesting application are
generalized-ensemble simulations of the random cluster model where one directly
estimates the number of possible graphs $g(k,b)$ with $k$ clusters and $b$ edges
which then allows for the calculation of canonical ensemble averages as continuous
functions of temperature and the parameter $q$.

The source code of the implementations discussed here, in particular including the
dynamic connectivity algorithm, is available on GitHub under a permissive
license \cite{sweenycode}.
 
\begin{acknowledgments}
  E.M.E. would like to thank T.\ Platini and N.\ Fytas for fruitful discussions and
  U.\ Wolff for providing an implementation of his automatic windowing method.
\end{acknowledgments}


%

\end{document}